\documentclass[pra,amsmath,preprintnumbers,amssymb,aps,superscriptaddress,twocolumn]{revtex4-2}

\usepackage[colorlinks=true,urlcolor=blue,citecolor=blue,allcolors=blue]{hyperref}

\usepackage{color}
\usepackage{braket}
\usepackage{amsmath}
\usepackage{esint}
\usepackage{amssymb}
\usepackage{physics}
\usepackage[utf8]{inputenc}
\usepackage{csquotes}
\usepackage{bm}
\usepackage{graphicx}
\usepackage{epstopdf}
\usepackage{tikz}
\usetikzlibrary{calc}
\usepackage{soul}
\usepackage{xcolor}
\usepackage{bookmark}
\usepackage{array}
\usepackage{slashed}
\usepackage{bbm}
\usepackage{microtype}
\usepackage{upgreek}
\usepackage[capitalize]{cleveref}

\newcommand{\MF}{\textrm{MF}}

\begin{document}
\title{Quantum optical impurity models in interacting waveguide QED}

\author{Adrian Paul Misselwitz}
\affiliation{Technical University of Munich, TUM School of Natural Sciences, Department of Physics, 85748 Garching, Germany}
\affiliation {Walther-Meißner-Institut, Bayerische Akademie der Wissenschaften, 85748 Garching, Germany}
\affiliation{Munich Center for Quantum Science and Technology (MCQST), 80799 Munich, Germany}

\author{Jacquelin Luneau}
\affiliation{Technical University of Munich, TUM School of Natural Sciences, Department of Physics, 85748 Garching, Germany}
\affiliation {Walther-Meißner-Institut, Bayerische Akademie der Wissenschaften, 85748 Garching, Germany}
\affiliation{Munich Center for Quantum Science and Technology (MCQST), 80799 Munich, Germany}

\author{Peter Rabl}
\affiliation{Technical University of Munich, TUM School of Natural Sciences, Department of Physics, 85748 Garching, Germany}
\affiliation {Walther-Meißner-Institut, Bayerische Akademie der Wissenschaften, 85748 Garching, Germany}
\affiliation{Munich Center for Quantum Science and Technology (MCQST), 80799 Munich, Germany}

\date{\today}

\begin{abstract}
We consider a generic model for interacting waveguide QED systems, where photons in a coupled-cavity array localize around atomic impurities while simultaneously interacting through local Kerr nonlinearities. This scenario appears naturally in nanophotonic crystals, circuit QED lattices, and ultracold atomic systems and is governed by the competition between attractive Jaynes-Cummings-mediated binding and intrinsic photon-photon repulsion. We analyze how this interplay affects the formation of localized few-photon bound states and determine the resulting many-body ground states for large periodic arrays of impurities and different filling factors. With the help of large-scale numerical simulations and approximate analytical models, we identify a rich phase diagram featuring Mott-like insulating states as well as superfluid phases with long-range correlations, which are mediated by an unbound, but strongly interacting photonic fluid.    
\end{abstract}

\maketitle
\section{Introduction}

The physics of impurities plays a crucial role in explaining many puzzling properties of solid-state systems and has thus been the subject of intense research for many decades.  A prominent example is the Kondo effect~\cite{kondo_resistance_1964}, which describes the unexpected increase in electrical resistance and the resulting anomalies in conduction of certain materials at very low temperatures. Within the Kondo model~\cite{Hewson_1993} or its extension, the Anderson impurity model~\cite{anderson_localized_1961,schrieffer_relation_1966}, this anomaly is explained by the formation of localized electron clouds that surround magnetic impurities. Even at the level of individual impurities, this effect is inherently nonperturbative and results in a non-trivial entangled state of the many-body electron system.

More recently, a very different class of impurity models has gained considerable attention in the field of waveguide QED~\cite{RoyRMP2017,ChangRMP2018,SheremetRMP2023},
where photons confined to one-dimensional (1D) or two-dimensional (2D) channels are strongly coupled to individual two-level atoms (TLAs) or other quantum emitters. In these systems, the propagating particles are bosonic in nature and interact with the TLAs via a Jaynes-Cummings (JC) coupling~\cite{larson_jaynescummings_2024}. In photonic-crystal waveguides~\cite{GobanNatComm2014,LodahlRMP2015} or coupled-cavity arrays~\cite{Lambropoulos2000}—where the group velocity is significantly reduced and the dispersion relation exhibits band gaps—this interaction can lead to the formation of atom-photon bound states~\cite{Bykov1975,John1994,Kofman1994}, where an emitted photon remains localized in the vicinity of the atom. While a single TLA can in principle bind an arbitrary number of photons~\cite{calajo_atom-field_2016, shi_bound_2016}, the inherent nonlinearity of the two-state system renders the exact determination of these multi-photon bound states an interesting and nontrivial problem. In periodic arrays, where each photonic lattice site is coupled to a TLA, the same JC nonlinearity can also prevent multiple photons from occupying the same site. This situation is described by the Jaynes-Cummings-Hubbard (JCH) model~\cite{hartmann_strongly_2006,greentree_quantum_2006,angelakis_photon-blockade-induced_2007,Rossini2007,Schmidt2009,koch2009SuperfluidMottInsulator,Tomadin2010,Hartmann_2016,noh_quantum_2017}, which predicts a quantum phase transition between a delocalized superfluid phase and an incompressible Mott-insulating photonic state. 

Current interest in such quantum-optical impurity models is driven by the rapid development of experimental platforms capable of realizing strongly coupled waveguide QED systems. These platforms include, for instance, nanophotonic crystal waveguides, where optical photons are coupled to atoms~\cite{HoodPNAS2016}, quantum dots~\cite{LodahlRMP2015} or defect centers~\cite{SipahigilScience2016}. Further, in the field of circuit QED~\cite{blais2021CircuitQuantumElectrodynamics}, artificial TLAs (i.e., superconducting qubits) can be coupled very efficiently to propagating microwave photons in transmission lines~\cite{AstafievScience2010} or coupled-resonator arrays~\cite{liu_quantum_2017,Ferreira2021,scigliuzzo_controlling_2022,JouannyNatComm2025}.  
Also, beyond photonic systems, it has been demonstrated~\cite{krinner_spontaneous_2018,kwon_formation_2022} that ultracold atoms in optical lattices can be used to realize analogous waveguide QED models, where the `photons' and `atoms' are represented by different internal states of bosonic atoms~\cite{deVegaPRL2008,Navarrete-Benlloch_2011}. As experimental control improves and systems scale toward larger 1D and 2D arrays, these platforms can serve as natural analogue simulators for exploring the intriguing few- and many-body quantum dynamics of bosonic impurity models.

In this work we go beyond the framework of conventional waveguide QED systems and consider 1D photonic arrays where---in addition to the coupling with TLAs---the photons interact among themselves via a strong local Kerr effect.  This generalization is motivated by the platforms mentioned above, where the strong transverse confinement of photons enhances intrinsic or engineered~\cite{hartmann_strongly_2006} material nonlinearities or nonlinearities stemming from the Josephson effect~\cite{ma2019DissipativelyStabilizedMott,saxberg2022DisorderassistedAssemblyStrongly,zhang2023SuperconductingQuantumSimulator}. In previous theoretical studies, it has already been shown that the presence of photon-photon interactions can drastically modify basic quantum optical effects, such as collective emission or photon-mediated dipole-dipole interactions~\cite{wang_supercorrelated_2020,TalukdarPRA2022,TalukdarPRA2023,roccati_many-body_2025} and give rise to new features that are not observed for atoms that are coupled to a purely linear photonic environment. 

Here we are interested in the few- and many-body correlations that emerge in the ground state of a lattice model for interacting waveguide QED. We first examine how the competition between attractive JC binding and photon-photon repulsion affects the formation of multi-photon bound states in this system and limits the maximal number of photons that can be attached to a single atomic impurity. By extending this analysis to large periodic arrays, we find that the remaining free photons can either be blocked or transmitted by the impurities, resulting in a rich phase diagram with Mott-insulating states and long-range correlated superfluid phases. Using both analytical approximations and large-scale numerical simulations, we evaluate the ground-state phase diagram for a fixed excitation density and discuss the relevant experimental quantities that distinguish the different phases. As an interesting outcome of this analysis, we find that the atom-photon coupling can serve as an effective chemical potential that controls the density of an otherwise isolated photonic many-body system. This property can be relevant for the simulation of a much broader class of isolated photonic many-body systems, where a chemical potential is not naturally available.

The remainder of the paper is structured as follows. In Sec.~\ref{sec:The_Model} we introduce the generic model for an  interacting waveguide QED system and use this model in Sec.~\ref{sec:one_atom} to analyze the bound states of a single atomic impurity. In Sec.~\ref{sec:periodic_system} we extend this analysis to periodic arrays and discuss the many-body ground-state phases of this system. Finally, in Sec.~\ref{sec:Implementations} we present two possible experimental implementations of our model and conclude our findings in Sec.~\ref{sec:Conclusions}.

\section{Interacting waveguide QED} \label{sec:The_Model}

\begin{figure}
\includegraphics[width=\linewidth]{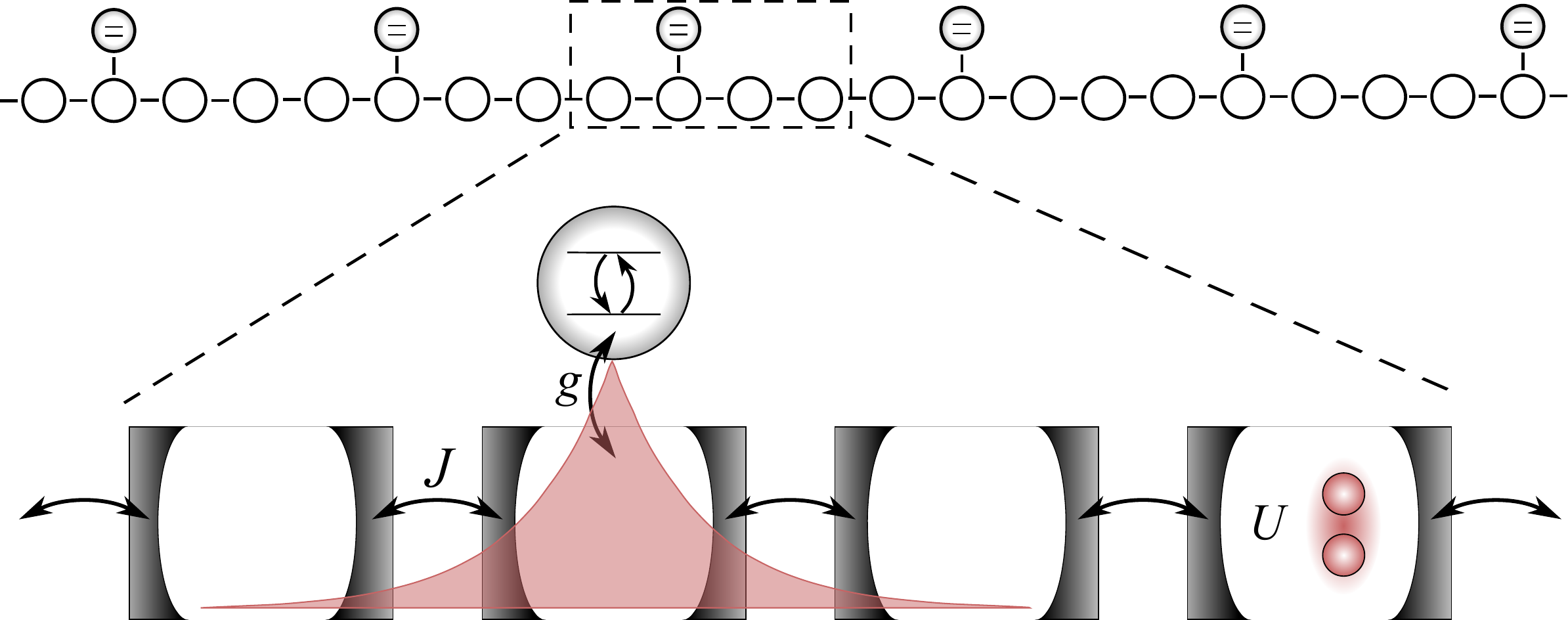}
\caption{Sketch of an interacting waveguide QED system. A set of TLAs with states $|g\rangle$ and $|e\rangle$ are coupled to photons propagating along an array of coupled cavities with nearest-neighbor hopping $J$. The TLAs couple to the local photonic mode with strength $g$ and photon-photon interaction within each cavity are modeled by Kerr-type process of strength $U$. In the considered periodic configurations, each unit cells highlighted by the dashed box contains one TLA and $d=L/N_{\rm a}$ cavities.}
\label{fig1:Setup}
\end{figure}

We consider a strongly interacting waveguide QED system as depicted in Fig.~\ref{fig1:Setup}, where a set of $N_{\rm a}$ TLAs or other quantum emitters is coupled to photons propagating along a 1D channel. The TLAs have a ground state $|g\rangle$ and an excited state $|e\rangle$, which are separated by a transition frequency $\omega_{\rm a}$. The waveguide is realized by an array of $L$ identical cavities, each representing a localized photonic mode with frequency $\omega_{\rm c}$ and bosonic annihilation (creation) operator $a_x$ ($a_x^\dag$), where $x=1,\dots,L$ labels the lattice site. A weak coupling between neighboring cavities with tunneling amplitude $J$ permits propagation of photons along the array, as described by a tight-binding model with a total width of the propagation band of $4J>0$.  

\subsection{Lattice Hamiltonian}
We assume that the $j$-th TLA couples to the local photonic mode at site $x_j$ via a JC interaction with a coupling constant $g$. In addition, the photons interact among themselves through a local, Kerr-type interaction of strength $U$. For periodic boundary conditions, the total Hamiltonian for this system is given by $(\hbar=1)$
\begin{equation}\label{eq:Hamiltonian}
\begin{split}
    H   = \,\, &  \sum_{x=1}^L \omega_{\rm c} a^{\dagger}_x a_{x} +\frac{U}{2} a^{\dagger}_x a^{\dagger}_xa_{x} a_{x}-  J  \left(a^{\dagger}_x a_{x+1} + {\rm H.c.} \right)\\
       & + \sum_{j=1}^{N_{\rm a}}  \omega_{\rm a}\sigma^{+}_j \sigma^{-}_j + g  \left(a^{\dagger}_{x_j} \sigma^{-}_j + a_{x_j} \sigma^{+}_j\right).
\end{split}
\end{equation} 
Note that here we have already assumed that $|J|,|g|,|U|\ll \omega_{\rm c}, \omega_{\rm a}$, such that only interactions that conserve the total number of excitations in the system are retained. Also, in the following analysis we will primarily focus on the case $U>0$, where the second term in Eq.~\eqref{eq:Hamiltonian} represents a repulsion between photons on the same lattice site. In Sec.~\ref{sec:Implementations} below we discuss in more detail how this model can be realized with coupled arrays of superconducting circuits and with ultracold atoms in state-dependent optical lattices.

\subsection{Relation to other models}
Equation~\eqref{eq:Hamiltonian} represents a general class of lattice Hamiltonians, from which various well-known models in the field of quantum optics, ultracold atoms, and condensed-matter physics are recovered in different limiting cases. For $J\rightarrow 0$, we obtain a set of independent JC models when $U=0$ or a set of independent Kerr oscillators when $g=0$. Both cases serve as minimal toy models for describing nonlinear quantum-optical effects at the level of individual photonic modes. For $J>0$ and $U=0$, one recovers the standard lattice Hamiltonian in waveguide QED~\cite{ZhouPRL2008,Hartmann_2016,noh_quantum_2017,calajo_atom-field_2016,shi_bound_2016}, where the local atom-photon coupling competes with the hopping-induced delocalization of the photons. For $g\gtrsim J$, photons localize around a single atomic impurity and form a so-called atom-photon bound state~\cite{Lambropoulos2000}. In a periodic system with one atom in each cavity,  we obtain the JCH Hamiltonian. For this model, a sharp quantum phase transition between a superfluid and a Mott-insulator phase has been predicted~\cite{greentree_quantum_2006,angelakis_photon-blockade-induced_2007,Rossini2007,Schmidt2009,koch2009SuperfluidMottInsulator,Tomadin2010,Hartmann_2016,noh_quantum_2017}, which arises from an effective photon-photon repulsion that is induced by the JC nonlinearity. A similar physics occurs for $g=0$ but $U>0$, where this repulsion arises from the direct  Kerr interaction between photons~\cite{hartmann_strongly_2006}. In this case, the photons are decoupled from the atoms and are described by the well-known Bose-Hubbard (BH) model~\cite{fisher1989BosonLocalizationSuperfluidinsulator}. 

Finally, in the single-impurity limit, our model relates to bosonic impurity models~\cite{lee_quantum_2007, lee_nrg_2010, warnes_symmetry_2012, foss-feig_phase_2011}, i.e., the bosonic counterparts of the fermionic Anderson-type models. Such models predict similar photonic localization effects as discussed below, but the models and the underlying mechanisms are not the same and there is no strict correspondence.

\subsection{From single impurities to periodic arrays}\label{subsec:Array}
Compared to the different limiting cases mentioned above, there is still little known about the general situation, where both Kerr-type and JC-type interactions are present at the same time. 
Here, we are interested in the ground-state properties of Hamiltonian $H$ in the limit $L\rightarrow\infty$, which, apart from the interaction parameters $g$ and $U$, will also depend crucially on the density of TLAs and the photonic filling factor. For concreteness, we restrict our analysis to waveguides with a periodic arrangement of TLAs and a fixed  total  number of excitations, 
\begin{equation} 
N_{\rm ex} =  \sum_{j=1}^{N_{\rm a}} \langle \sigma_j^{+} \sigma^{-}_j\rangle + \sum_{x=1}^L \langle a^{\dagger}_x a_{x}\rangle,
\end{equation} 
which is conserved under the dynamics of $H$. We denote by $d=L/N_{\rm a}$ the distance between two TLAs, such that each unit cell consists of $d$ cavities and contains on average $N=N_{\rm ex}d/L$ excitations. For $d\gg1$, we then recover the single-impurity limit where the physics of this system can be understood in terms of individual impurity atoms that localize photons around them. For $d\sim  O(1)$, photons bound to neighboring TLAs start to overlap~\cite{calajo_atom-field_2016,shi_effective_2018} and interact with unbound photons in between impurities. In this case, nontrivial correlations can be established even across distant unit cells.

\section{Interacting photons bound to a single impurity atom} \label{sec:one_atom}

In this section, we first consider the case of a single TLA that is coupled to an infinite photonic waveguide,  i.e., $d=L\rightarrow \infty$. For $U=0$ it has been previously shown that in this scenario photons can lower their energy by binding to the atom. This effect leads to the formation of so-called atom-photon bound states, with single~\cite{Bykov1975,John1994,Kofman1994} or multiple~\cite{calajo_atom-field_2016, shi_bound_2016} photons localized around the atomic impurity. For finite $U>0$, the repulsion between the photons counteracts this tendency and depending on the values of $U$, $g$ and $J$, we expect that only a finite number of photons can be bound to a single atom.  

\subsection{Atom-photon bound states}
In the limit $U\rightarrow 0$ and for a single excitation, $N=1$, the eigenstates of Hamiltonian $H$ can be solved exactly. They can be grouped into scattering states $|k\rangle$, which are extended over the whole lattice and form a band of energies $E_k=\omega_{\rm c}-2J\cos(k)$ with $k\in[-\pi,\pi]$. In addition, there are two bound eigenstates $\ket{n=1,\pm}$ with energies $E_{n=1,\pm}$ that lie outside this propagation band. For the resonant case, $\Delta=\omega_{\rm c}-\omega_{\rm a}=0$, we obtain~\cite{calajo_atom-field_2016}
\begin{equation}\label{eq:E_1bound_anyJ}
E_{1,\pm}= \omega_c\pm \sqrt{2J^2 + \sqrt{4J^4+g^4}},
\end{equation}
and the corresponding eigenstates can be written as 
\begin{equation}
    \ket{1,\pm} = \left[ \cos \theta_\pm \sigma_+ \pm \sin \theta_\pm a_{\lambda, \pm}^{\dagger}(x_{\rm a}) \right] \ket{g,0}.
\end{equation}
Here, the mixing angle $\theta_{\pm}$ determines whether the bound state is more atom-like or more photon-like and 
\begin{equation}
    a_{\lambda, \pm}^{\dagger}(x_{\rm a}) = \sum_x \frac{(\mp)^{|x-x_{\rm a}|} e^{-\frac{|x-x_{\rm a}|}{\lambda_\pm}}}{\sqrt{\coth (1/\lambda_\pm)}} a_x^{\dagger}
\end{equation}
is a bosonic creation operator. It creates a photon that is exponentially localized 
around the atom at site $x_{\rm a}$. The localization length, $\lambda_{\pm}$,  is finite for any $g$, but diverges as $\lambda_\pm\simeq 4J^2/g^2$ in the limit $g/J\rightarrow 0$. 

\begin{figure}
\includegraphics[width=\linewidth]{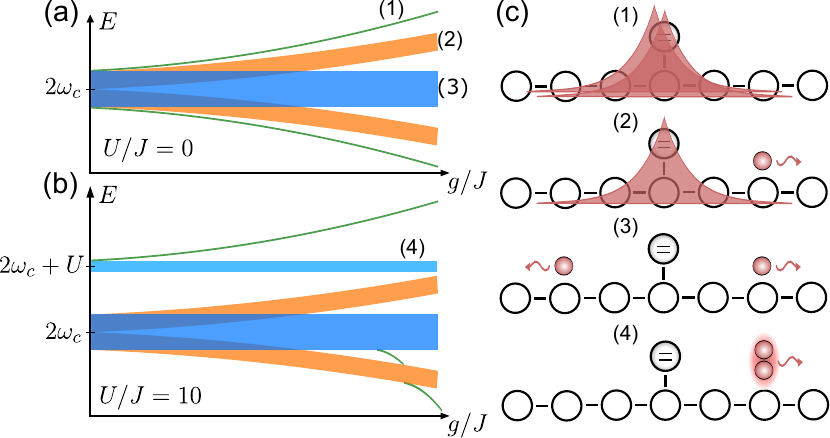}
\caption{Sketch of the full spectrum of the waveguide Hamiltonian given in Eq.~\eqref{eq:Hamiltonian} for $N_{\rm a}=1$ TLA and $N=2$ excitations. 
The energies are plotted as a function of $g/J$ and for the cases (a) $U=0$ and (b) $U/J=10$. The colors and labels indicate the different types of energy eigenstates, which are also illustrated in (c). These are: (1) localized two-photon bound states (green line), (2) states with one bound and one free photon (orange), (3) states with two free photons (dark blue) and (4) propagating repulsively bound photon pairs (light blue). See text for more details.
}
\label{fig:ED_Sketch}
\end{figure}

Beyond the single excitation subspace, $N>1$, no simple analytic solutions for the $N$-photon bound state energies $E_{n=N,\pm}$ exist, but variational and numerical methods confirm the existence of bound states also for larger photon numbers~\cite{calajo_atom-field_2016, shi_bound_2016}. In this case, however, the localization weakens with increasing $N$. In the single-cavity limit, $g/J\rightarrow \infty$, the $n$-photon bound state energies approach the values
\begin{equation}\label{eq:JCsplitting}
E_{n,\pm}\approx n \omega_{\rm c} \pm g \sqrt{n}, 
\end{equation}
which are the usual dressed-state energies of the JC model. This expressions shows that in the ground state and for large $g$, all photons are bound to the atomic impurity. However, the binding energy per photon decreases with the total number of photons. For $N=2$ the full energy spectrum of $H$ is plotted as a function of $g/J$ in Fig.~\ref{fig:ED_Sketch}(a).

\subsection{Bound states of interacting photons: single-cavity model}\label{subsec:JCmodel}
In the presence of a Kerr nonlinearity, there is an additional energy cost of $Un(n-1)/2$ for $n$ photons occupying the same cavity mode. For $U<0$ this effect simply increases the binding of the photons to the atom. However, when compared to Eq.~\eqref{eq:JCsplitting}, we see that for the case of interest, $U>0$, this repulsion eventually exceeds the JC binding energy, when the photon number is large enough. This is intuition is confirmed by the exact spectrum shown in Fig.~\ref{fig:ED_Sketch}(b) for $N=2$.

\begin{figure}
\includegraphics[width=\linewidth]{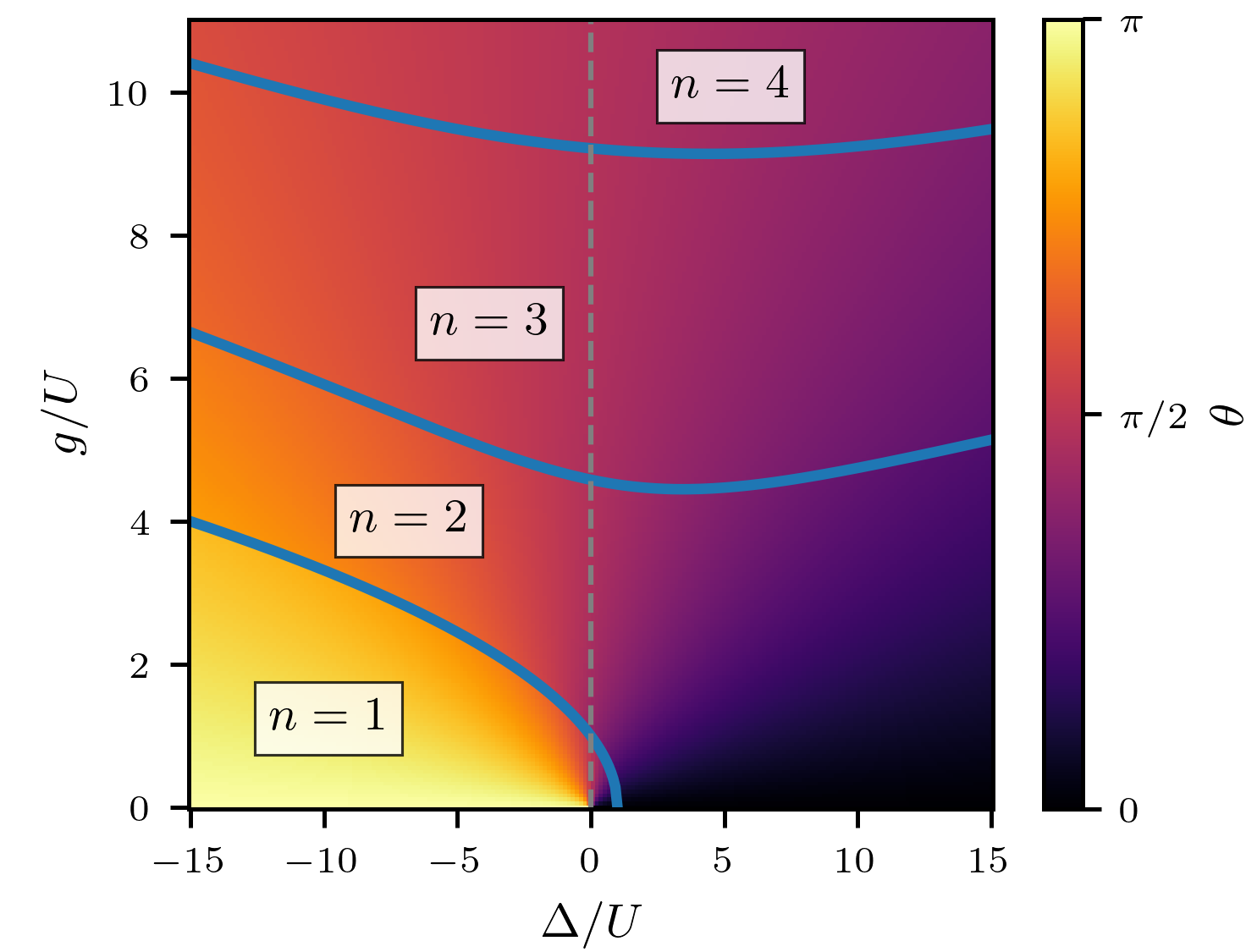}
\caption{Plot of the maximal number of photons that can be bound to a single TLA in the limit $J\rightarrow 0$. The boundaries between the different domains (blue lines) are determined by the minimal binding strength $g_{\rm b}(n)$, evaluated from Eq.~\eqref{eq:U_critical_condition}. The color indicates the value of the mixing angle $\theta_n$ of the corresponding ground state
$\ket{n,-} = \sin(\theta_n/2)\ket{g,n} - \cos(\theta_n/2)\ket{e,n-1}$, which is more `atom-like' when $\theta_n\approx 0$ and more `photon-like' when $\theta_n\approx \pi$.
}
\label{fig:transitions_J0_limit}
\end{figure}

To analyze this scenario, it is instructive to consider first the limit $J\rightarrow 0$ and study the resulting single-site JC model with an additional Kerr nonlinearity. This model conserves the total number of excitations and can be diagonalized within each two-state subspace $\{\ket{g,n}, \ket{e,n-1}\}$ separately. The resulting eigenstates and eigenenergies are the same as in the regular JC model~\cite{larson_jaynescummings_2024}, but with the substitution 
\begin{equation}
    \Delta \rightarrow \Delta_n = \Delta + U(n-1). 
\end{equation}
Therefore, we obtain the eigenstates 
\begin{align}
    \ket{n,+} &= \cos(\frac{\theta_n}{2})\ket{g,n} + \sin(\frac{\theta_n}{2})\ket{e,n-1}, \\
    \ket{n,-} &= \sin(\frac{\theta_n}{2})\ket{g,n} - \cos(\frac{\theta_n}{2})\ket{e,n-1},
    \label{eq:JC_ev}
\end{align}
with a mixing angle that satisfies
\begin{equation}
    \tan(\theta_n) = \frac{2g\sqrt{n}}{\Delta_n} = \frac{2g\sqrt{n}}{\Delta + U(n-1)}.
\end{equation}
The corresponding energies are
\begin{equation}
\begin{split}
    E_{n,\pm}^{\rm JC} =\, &n\omega_{\rm c}  - \frac{\Delta}{2}+\frac{U}{2}(n-1)^2\\
    &\pm\frac{1}{2}\sqrt{[\Delta+U(n-1)]^2+4ng^2}.
    \label{eq:JC_energies}
\end{split}
\end{equation}
This result already allows us to establish a simple minimal condition for the existence of an $n$-photon bound state, namely 
\begin{align}\label{eq:U_critical_condition}
    E_{n,-}^{\rm JC} \leq E_{n-1,-}^{\rm JC} + \omega_{\rm c}.
\end{align}
If this condition is not satisfied, it becomes energetically more favorable to remove one photon from the coupled atom-cavity site and place it somewhere else in the waveguide. Therefore, in the limit $J\rightarrow0$, the boundary set by  Eq.~\eqref{eq:U_critical_condition} determines the minimal coupling strength $g_{\rm b}(n)$, above which there are at least $n$ photons bound to the atom in the ground state. 
For $\Delta=0$ we obtain 
\begin{align}\label{eq:g_critical_0detuning}
    g_{\rm b}(n)/U = \sqrt{(2n-3)[2n(n-2)+1]}.
\end{align}
For nonvanishing detunings, the domains with different numbers of bound photons are plotted in Fig.~\ref{fig:transitions_J0_limit} together with the value of the corresponding mixing angle $\theta_n$.

Note that for $\Delta/U>1$, i.e., when $\omega_{\rm c}>\omega_{\rm a}+U$, a two-photon bound exists for arbitrary small $g\ll U$.
This can be understood by considering the eigenstates in the limit $g\ll U,\Delta$.
For $g\rightarrow 0$ and $\Delta>0$, the $n=1$ and $n=2$ ground states of the JC system are $\ket{e,0}$ and $\ket{e,1}$ with energies $E_{1,-}^{\rm JC}=\omega_c-\Delta$ and $E_{2,-}^{\rm JC}=2\omega_c-\Delta$, respectively. 
For small $g$, these states couple to the higher energy states $\ket{g,1}$ and $\ket{g,2}$ with the respective couplings $g$ and $\sqrt{2}g$.
This results in a second-order correction to the ground state energy of $\delta E^{(2)}_{1,-}=-g^2/\Delta$ and $\delta E^{(2)}_{2,-}=-2g^2/(\Delta+U)$, such that according to Eq.~\eqref{eq:U_critical_condition} the two photon bound state is favored for any $\Delta>U$.

\subsection{Photon capture and detachment transitions} \label{sec:Groundstate_transition}
Near the coupling $g_{\rm b}(n)$, the $n$-photon bound state and any state with $n-1$ bound photons and one photon in the waveguide are nearly degenerate and hybridize for nonvanishing $J$. To capture this transition region more accurately, we consider a system of $N$ excitations in total and---assuming $g\approx g_{\rm b}(N)$---include the hopping of photons in the waveguide using degenerate perturbation theory. 

As discussed above, exactly at the transition and for $J=0$, we have a degeneracy between the states
\begin{equation}
    \ket{\psi_0} = P^\dagger_{N}\ket{\text{vac}}
    \quad\text{and}
    \quad
    \ket{\psi_{x\neq 0}} \equiv P_{N-1}^\dagger a_x^\dagger \ket{\text{vac}},
    \label{eq:polariton_def}
\end{equation}
where $\ket{\text{vac}}$ is the vacuum state and  the operator $P_n^\dagger = \ket{n,-}_0\bra{g,0}$ creates an $n$-photon polariton with energy $E_{n,-}$ at site $x_{\rm a}=0$.
For $g$ close to $g_{\rm b}(N)$, the effective Hamiltonian at first order in $J$ is $H_\text{eff} = \mathbbm{P}H\mathbbm{P}$ with $\mathbbm{P}=\sum_x \ket{\psi_x}\bra{\psi_x}$.
The perturbation doesn't change the on-site energy and the hopping between the cavities uncoupled to the atom. Therefore, for $x \neq 0$, 
\begin{align}
    \bra{\psi_x}H_\text{eff}\ket{\psi_x} = E_{N-1,-}^{\rm JC} + \omega_{\rm c}.
\end{align}
For $x \neq 0$ and $x \neq -1$ we obtain 
\begin{align}
    \bra{\psi_0}H_\text{eff}\ket{\psi_0} &= E_{N,-}^{\rm JC} = E_{N-1,-}^{\rm JC} + \omega_{\rm c} + \epsilon_N,\\
    \bra{\psi_{x+1}}H_\text{eff}\ket{\psi_x} &= -J,
\end{align}
with a local energy offset of
\begin{equation}
    \epsilon_N = E_{N,-}^{\rm JC} - E_{N-1,-}^{\rm JC}-\omega_{\rm c}.
\end{equation}
In addition, the hopping amplitude between the impurity site and the neighboring cavities is modified as
\begin{align}
    \bra{\psi_{\pm 1}}H_\text{eff}\ket{\psi_{0}} = - \alpha_N J,
\end{align}
where
\begin{align}\label{eq:alpha}
    \alpha_N =& \bra{-,N-1}a_0\ket{-,N}\nonumber \\
    =& \sqrt{N}\sin(\frac{\theta_{N-1}}{2})\sin(\frac{\theta_{N}}{2}) \nonumber\\
    &+ \sqrt{N-1}\cos(\frac{\theta_{N-1}}{2})\cos(\frac{\theta_{N}}{2}).
\end{align}
Therefore, the effective Hamiltonian $H_\text{eff}$ describes a single-particle tight-binding model with an impurity that is separated from the remaining lattice sites by an  energy offset $\epsilon_N$ and a modified tunneling matrix element $\alpha_N J$.   

\begin{figure}
    \centering
    \includegraphics[width=\linewidth]{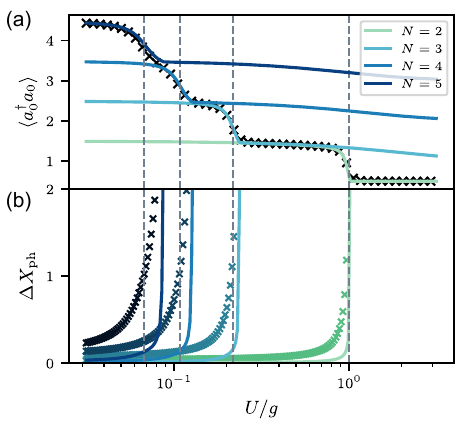}
    \caption{Detachment of bound photons. (a) Plot of the number of bound photons at the impurity site, $\langle a_0^\dag a_0\rangle$, in the ground state of $H$ with $N_{\rm a}=1$ TLA and a total number of $N=5$ excitations. The exact numerical results (crosses) are compared with the predictions of Eq.~\eqref{eq:cav_occ_pertbJ} for the effective model $H_{\rm eff}$ for different $N$. The same comparison is shown in (b) for the width of the localized photon wavepacket, $\Delta X_{\rm ph}$, as defined in Eq.~\eqref{eq:Xph}. For this comparison, also the exact numerics is performed for different excitation numbers to avoid contributions from the remaining unbound photons. In both plots, the vertical gray dashed lines indicate the value of $U/g_{\rm  b}(N)$ predicted from the single-site JC model in Sec.~\ref{subsec:JCmodel}. For all simulations, we have assumed the same value of $J/g=0.01$ and used $L=13$ ($L=100$) for the exact (effective) model.
    }
   \label{fig:Occupation_singleAtom_perturbJ}
\end{figure}

The ground state of the effective Hamiltonian can be decomposed as $\ket{\psi_{\rm GS}}=\sum_x c_x \ket{\psi_x}$, an solved analytically, for example by adapting the methods from Ref.~\cite{pury_mixed_1991}. Here we do not carry out this calculation explicitly, but focus instead on a numerical comparison between $\ket{\psi_{\rm GS}}$ and the exact ground state of the full lattice Hamiltonian. To do so we evaluate the photon number of the impurity site, 
\begin{align}
    &\bra{\psi_{\rm GS}}a^\dagger_0 a_0\ket{\psi_{\rm GS}} =\sum_{x}|c_x|^2 \bra{\psi_x}a^\dagger_0 a_0\ket{\psi_x} 
   \label{eq:cav_occ_pertbJ}
\end{align}
together with the width of the photonic wavefunction
\begin{equation}\label{eq:Xph}
\Delta X_{\rm ph} =  \sqrt{ \bra{\psi_{\rm GS}}X_{\rm ph}^2 \ket{\psi_{\rm GS}}  }, 
\end{equation}
where $X_{\rm ph}=\sum_x x \,a^\dag_x a_x$. These quantities are plotted in Fig.~\ref{fig:Occupation_singleAtom_perturbJ} for a small, but nonvanishing $J$ and compared to the corresponding quantities obtained from exact diagonalization of the full Hamiltonian $H$ for different $N$. We see that for these parameters, the effective model captures very well the successive detachment of individual bound photon as the photon-photon repulsion $U$ is increased. For small $N$ the transition points agree well with the analytic predictions for $g_{\rm b}(N)$ given in Eq.~\eqref{eq:g_critical_0detuning}. For larger $N$, slightly larger values of $U$ are allowed before photons detach. 
We attribute this shift to the effective hopping impurity~$\alpha_N J$, which contributes to the photon binding, such that a finite energy offset $\epsilon_N>0$ is required to detach the photon.
In addition, the finite extension of the multi-photon bound state reduces the impact of interactions compared to the single-cavity model. 
Similarly, the effective model underestimates the width of the multi-photon wavepacket, as it treats the $N-1$ bound photons as being localized on a single site.

\subsection{Binding threshold for weakly bound photons}
So far we have focused on the strong coupling regime $g\gg J$, where the binding threshold for an $N$-photon bound state is determined by the competition between local coupling and interaction processes. In the opposite regime, $g\ll J$, it is known that atom-photon bound states extend over many lattice sites, $\lambda_-\gg1$, which also reduces the impact of photon-photon repulsion. 

\begin{figure}
\centering
\includegraphics[width=\linewidth]{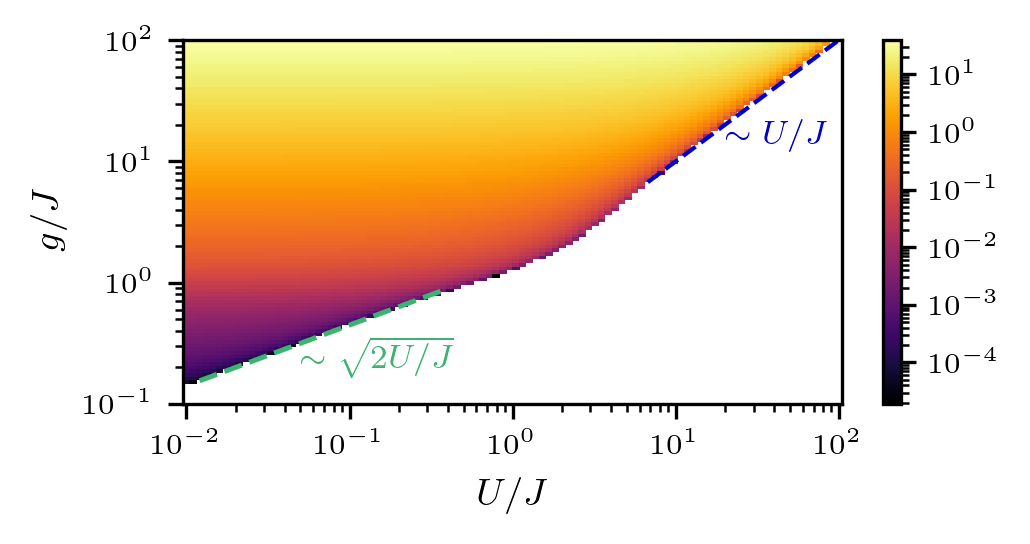}
\caption{Plot of the binding energy $E_{\rm b}(N=2)$ introduced in Eq.~\eqref{eq:energy_difference} for different JC couplings and interaction strengths and for $\Delta=0$. The dashed lines indicate the analytic expressions for $g_{\rm b}(N=2)$, as obtained from Eq.~\eqref{eq:g_critical_0detuning} and Eq.~\eqref{eq:gb_weak} in the strong-coupling and in the weak-coupling limit, respectively. For simulations we have used a system of size $L=40$ and replaced $E_{1,-}\rightarrow E_{1,-}+ U \cos{\theta_1}/L$ in the definition of $E_{\rm b}(N=2)$. This adaption is necessary to correctly account for the residual interaction between the unbound photon and the impurity site for any finite $L$. 
}
\label{fig:Energy_Difference}
\end{figure}

To study the existence of  multi-photon bound states in this weak-coupling regime, we perform exact diagonalization of the full lattice Hamiltonian for $N_{\rm a}=1$ and $N=2$ excitations.  For this setting, the ground state energy, $E_{\rm GS}$, is evaluated for $L\gg1$ and used to compute the two-photon binding energy
\begin{equation}\label{eq:energy_difference}
E_{\rm b}(N=2)= {\rm min}\{E_{1,-}+\omega_{\rm c}-2J -E_{\rm GS}, 0 \}.
\end{equation}
This binding energy is plotted in Fig.~\ref{fig:Energy_Difference} for different ratios of $U/J$ and $g/J$ and for $\Delta=0$. For small $J$, we see that the threshold $E_{\rm b}(N=2)=0$ agrees well with the prediction $g_{\rm b}(2)=U$ from  Eq.~\eqref{eq:g_critical_0detuning}. For larger values of $J$, there is still a minimal coupling strength $g_{\rm b}(2)$ for binding two photons, however, with a different dependence on $U$.

To explain the scaling of the two-photon binding threshold in the weak-coupling regime, we make the simplifying assumption that for $U=0$, both photons are exponentially localized around the TLA with a length $\lambda_-$ given by the single-photon bound state. This is not exact, but a reasonably accurate assumption~\cite{calajo_atom-field_2016}. In the limit~$J\gg g$, we obtain $\lambda_-\simeq 4J^2/g^2$ and a single-photon bound state energy of
\begin{align}
    E_{1,-} \simeq  \omega_{\rm c}-2J-\frac{g^4}{16J^3}.
\end{align}
For finite $U$, the relevant interaction energy is given by 
\begin{align}
    E_{2,U}\simeq  \frac{U}{\coth^2(1/\lambda_-)}\sum_x e^{-4|x|/\lambda_-}\simeq\frac{U}{2\lambda_-}.
\end{align}
Thus, we can estimate the minimal coupling strength $g_{\rm b}(N=2)$ to bind two photons by setting
\begin{align}
    E_{1,-}+\omega_c-2J = E_{2,-}\simeq 2E_{1,-} + E_{2,U}.
\end{align}
Since in the weak-coupling limit $E_U\simeq Ug^2/(8J^2)$, we obtain the simple expression
\begin{align}
   g_{\rm b}(N=2)\simeq \sqrt{2U J},
   \label{eq:gb_weak}
\end{align}
which matches very accurately the numerical results in this limit.

For $N>2$, the same approximations can be used to estimate the energy of a weakly bound $N$-photon state,
\begin{align}
    E_{N,-}\simeq NE_{1,-}+E_{N,U},
\end{align}
where the interaction energy satisfies $E_{N,U}=E_{N-1,U}+(N-1)E_{2,U}$.
The $N$ photons are bound to the TLA when 
$    E_{N,-}\leq E_{N-1,-}+\omega_c-2J$,
which translates into a binding threshold of 
\begin{align}
    g_{\rm b}(N)\simeq \sqrt{2(N-1)UJ}.
\end{align}

\subsection{Correlations between unbound photons}
\label{sec:confined_photons}
\begin{figure}
    \centering
    \includegraphics[width=\linewidth]{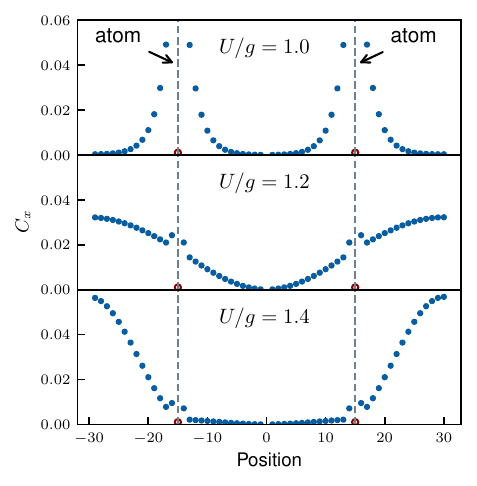}
    \caption{Correlations between unbound photons. The plots show the photon-photon correlation function $C_x=\langle  a^\dag_xa_x a^\dag_0 a_0 \rangle/\langle (a^\dag_0 a_0)^2\rangle$ evaluated for a periodic waveguide with $L=60$ lattices sites and $N_{\rm a}=2$ TLAs located at $x_{1,2}=\pm15$. The correlations are evaluated for a system with a total number of $N=4$ excitations, a hopping amplitude of $J/g=0.1$ and for different ratios of $U/g$. Note that values of $C_{x=\pm 15}$, which are strongly affected by the remaining bound photons, have been excluded from all plots. 
    }
    \label{fig:localized_photons}
\end{figure}

For a single TLA and a sufficiently large system, those photons which are not bound to the atom simply propagate freely through the waveguide. However, for systems with multiple impurities and finite excitation  density, these unbound photons interact with neighboring impurity sites and also among themselves. By that, they can establish long-range correlations across many impurity sites. To investigate the behavior of the unbound photons in a simple setting, we consider a small lattice with $L$ sites, periodic boundary conditions and two TLAs that are located at sites $x_{1,2}=\pm L/4$. For this system we use exact diagonalization to evaluate the ground state of the sector with a total number of $N_{\rm ex}=4$ excitations, such that for $g/U\gg 1$ each TLA binds $N=2$ photons. In Fig.~\ref{fig:localized_photons}, we then plot the normalized photon-photon correlations $C_x=\langle  a^\dag_xa_x a^\dag_0 a_0 \rangle/\langle (a^\dag_0 a_0)^2\rangle$ for different ratios of $U/g\in [1,1.5]$. Note that the site $x=0$ is located right in between the two TLAs such that for any $x\neq x_{1,2}$, this quantity mainly captures correlations between the unbound photons.

For $U/g=1$, despite the absence of an energy offset in the effective model $H_{\rm eff}$, i.e., $\epsilon_2=0$, two photons can still be bound to the atoms when $J/g=0.1$ due to the hopping impurity with~$\alpha_2\simeq 1.15$. Therefore, as seen in Fig.~\ref{fig:localized_photons}(a), the correlations for this set of parameters are strongly peaked around the locations of the atoms.  
For larger $U$, two photos detach from the TLAs and can propagate through the waveguide. Here we can distinguish two situations.
For~$U/g\simeq 1.2$, as shown in Fig.~\ref{fig:localized_photons}(b), the photons are anti-correlated with a profile that matches that of two repulsive bosons in a homogeneous lattice. In this regime, the photons are barely affected by the remaining atom-photon bound state. This can also be understood from the effective model $H_{\rm eff}$, where the corresponding parameters $\alpha_2\simeq 1.14$ and $\epsilon_2/g\simeq 0.06$ result in a transparent impurity, and the ground state of $H_{\rm eff}$ is a homogeneous plane wave.

When $U$ increases further, a qualitatively different picture arises. The unbound photons are still anticorrelated, but now $C_x\simeq 0$ across the whole region between the TLAs [see Fig.~\ref{fig:localized_photons}(c)]. This observation implies that the atomic impurities now act as impenetrable barriers for the unbound photons, which remain localized on opposite sites of those effective potential walls. Again, this picture is consistent with the effective model $H_{\rm eff}$, where, for $U/g=1.4$ and $J/g=0.1$, the parameters $\alpha_2\simeq1.13$ and $\epsilon_2/g\simeq 0.12$  correspond to a well-separated impurity site.
The photon repulsion then localizes one photon in each of the separated domains.
Below we find that this difference between unbound photons being simply trapped in between impurity sites or correlated across impurities also leads to a qualitatively different scaling of correlations in larger systems.

\section{Many-body correlations in periodic arrays} \label{sec:periodic_system}

In a next step, we go beyond the single-impurity limit and focus on the ground-state phases of an extended waveguide system with regularly spaced atomic impurities. As already discussed in Sec.~\ref{subsec:Array}, we do so by considering a fixed total number of $N$ excitations per until cell and $d$ is the distance between the equally-spaced TLAs. For the study of ground states, we exclusively focus on repulsive interactions $U>0$. The reason is that in the opposite, attractive case, it becomes energetically most favorable to localize all photons on the same lattice site~\cite{dur_three_2000} and for a fixed density there is no meaningful thermodynamic limit. This regime, however, could be probed in quench dynamics or using quasi-adiabatic preparation schemes, where such extreme states do not appear on experimentally relevant timescales.

\subsection{Effective polariton model}
In the previous section we have already shown that in the limit $g/J\gg1$ the low-energy physics of a single impurity can be well-described by strongly localized atom-photon bound states and freely propagating photons in the waveguide. By assuming that in the low energy subspace, this picture also holds for multiple impurities and moderate $g/J$, we can approximate $H$ in Eq.~\eqref{eq:Hamiltonian} by an effective polariton Hamiltonian of the form
\begin{align}\label{eq:H_polariton}
    H_{\rm pol} =& \sum_{x}E_{x}(n_x) \\
    &-J\sum_{x}\left[b_{x+1}^\dagger K_{x+1}(n_{x+1})K_x(n_x)b_x+ {\rm H.c.}\right]. \nonumber
\end{align}
Here, we have introduced a site-dependent set of polariton annihilation operators 
\begin{align}
    b_x &= \sum_n \sqrt{n}\ket{n-1,-}_x\bra{n,-} \quad&\text{for }x\in A, \\
    b_x &= a_x \quad&\text{for }x\notin A,
\end{align}
together with the number operators $n_x=b_x^\dagger b_x$,
where $A=\{x_j ~|~j=1,\dots,N_{\rm a}\}$ is the subset of lattice sites that contain a TLA. The corresponding onsite energies are 
\begin{align}
    E_x(n)&=E_{n,-}^{\rm JC}\quad&\text{for }x\in A,\\
    E_x(n)&=\omega_c n + \frac{U}{2}n(n-1)\quad&\text{for }x\notin A,
\end{align} 
and
\begin{align}
    K_x(n)&=\frac{\alpha_{n+1}}{\sqrt{n+1}}\quad&\text{for }x\in A,\\
    K_x(n)&=1\quad&\text{for }x\notin A,
\end{align}
account for the modified hopping matrix elements whenever a photon is hopping onto or off an impurity site.

\begin{figure}
    \centering
    \includegraphics[width=\linewidth]{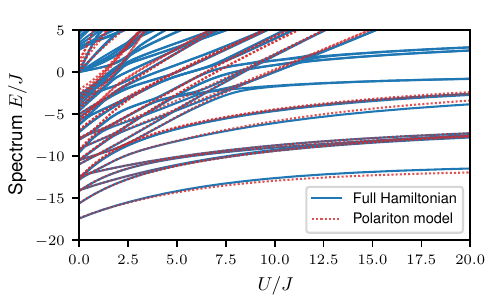}
    \caption{Comparison of the spectrum of the effective polariton model in Eq.~\eqref{eq:H_polariton} (red dashed lines) with the spectrum of the full waveguide Hamiltonian. For both models, the spectra are obtained using exact diagonalization for a small lattice of $L=4$ sites, $d=2$ and $N=4$ excitations. The JC coupling is set to $g/J=5$.}
    \label{fig:LLE_comparison_full}
\end{figure}

The Hamiltonian $H_{\rm pol}$ given in Eq.~\eqref{eq:H_polariton} generalizes the single-excitation model $H_{\rm eff}$ introduced in Sec.~\ref{sec:Groundstate_transition} to the many-body case. While it still contains most of the complexity of the original model, it provides additional intuition in terms of the bound polariton states and can be used as a starting point for further approximations. Although $H_{\rm pol}$ ignores the finite extent of the bound photons and is strictly valid only in the limit $J\rightarrow 0$, a comparison with the spectrum of the full model for a small lattice shown in Fig.~\ref{fig:LLE_comparison_full} confirms that it also captures very accurately the low-energy physics for moderate $J/g$. It is thus suited to model most of the transitions between Mott insulating and superfluid phases that we explore in the following.

\subsection{The Mott-superfluid phase transition in waveguides with non-interacting photons}

\begin{figure*}
    \centering
    \includegraphics[width= \linewidth]{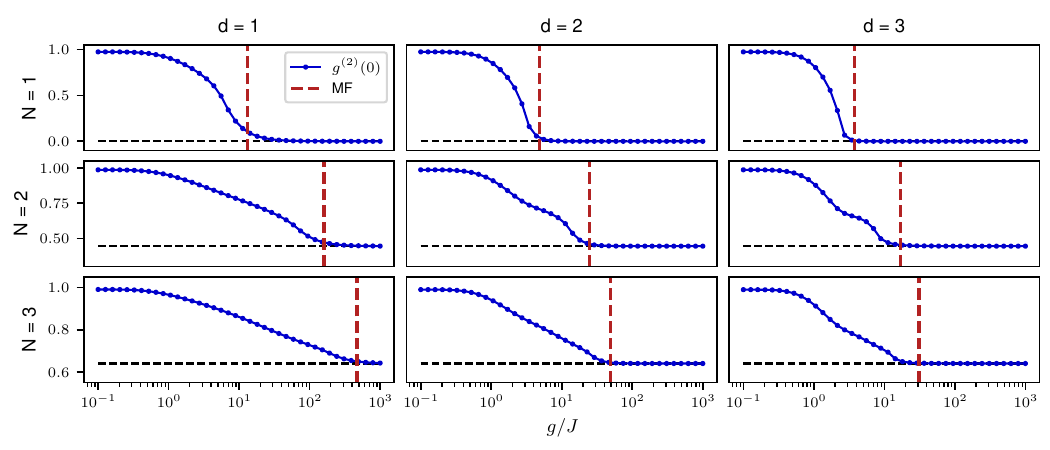}
    \caption{
    Ground-state phases of the JCH model with $U=0$. The phases are characterized in terms of the two-photon correlation function $g^{(2)}(0)$ defined in Eq.~\eqref{eq:g2} and averaged over all impurity sites. The individual plots show the dependence of $g^{(2)}(0)$ on the ratio $g/J$ for different excitation numbers $N$ (rows) and different impurity spacings (columns). The horizontal black dashed lines indicate the values of $g^{(2)}(0)$ given in Eq.~\eqref{eq:g2analytic}. The red vertical dashed lines mark the position of the critical coupling strength obtained from the mean-field analysis presented in Appendix~\ref{app:mean_field}. All numerical simulations have been performed for a lattice of $40$ unit cells, a maximal bond dimension of $\chi = 200$ and a truncation of the local Hilbertspace at $n_{\rm trunc} = 20$ photons.
    }
    \label{fig:var_sweep} 
\end{figure*}

In a first step, we revisit the case $U=0$, where our model reduces to the JCH model for $d=1$. In this case, the ground-state phases of this model have been analyzed previously using mean-field theory,  exact diagonalization and other numerical methods~\cite{hartmann_strongly_2006,greentree_quantum_2006,angelakis_photon-blockade-induced_2007,Rossini2007,Schmidt2009,koch2009SuperfluidMottInsulator,Tomadin2010,Hartmann_2016,noh_quantum_2017}. In these studies one finds a transition between a Mott insulator state for $g/J\gg1 $ and a superfluid state with long-range correlations in the opposite regime. The transition between the two phases is usually studied for a variable photon density and as a function of an additional chemical potential $\mu$. As a result one finds a phase diagram that displays characteristic Mott lobes, which can be distinguished from the superfluid phase in terms of a vanishing compressibility, $\kappa=\partial \langle a_x^\dag a_x\rangle/\partial \mu=0$.
 
Here we take a different approach and evaluate an experimentally more relevant scenario, where the total number of excitations in the lattice, $N_{\rm ex}$, is fixed. For this case we use large-scale numerical simulations based on a matrix-product-state ansatz (see Appendix~\ref{app:Numerics} for more details) to evaluate the ground-state phases of this model as a function of the ratio $g/J$ and different impurity separations $d$.  To distinguish different phases, we calculate the equal-time photon-photon correlation function 
\begin{equation}
g^{(2)}(0) = \frac{\langle a^\dag a^\dag a a \rangle }{\langle a^\dag a\rangle^2}.
\label{eq:g2}
\end{equation}
This quantity assumes a value of $g^{(2)}(0)\simeq 1$, for a coherent state and $g^{(2)}(0)\simeq 1-1/n$ for an $n$-photon  number state. Compared to the superfluid order parameter $\langle a\rangle$, the evaluation of $g^{(2)}(0)$ does not require an explicit breaking of the $U(1)$ symmetry of the model, and the quantity is experimentally accessible through photon-correlation measurements.

In the first column of Fig.~\ref{fig:var_sweep}, we show, first of all, the variation of $g^{(2)}(0)$ as a function of $g/J$ for $d=1$ and for varying excitation densities. These results reproduce the expected transition from a superfluid into a Mott insulating phase, as $g/J$ is increased. Note that the scaling $\sim \sqrt{n}$ of the JC energies ensures that deep in the Mott phase, all bound photons are evenly distributed among the TLAs, instead of being localized at a single impurity site. This corresponds approximately to a product state of the form $\prod_x (\ket{g,N}_x-\ket{e,N-1}_x)/\sqrt{2}$. This phase is characterized by a strongly antibunched value of
\begin{equation}\label{eq:g2analytic}
g^{(2)}(0) = \frac{(N-1)^2}{(N-\frac{1}{2})^2},
\end{equation}
when evaluated for one of the impurity sites. In the superfluid phase, the same correlation function increases and reaches a value of 
$g^{(2)}(0) \simeq 1$ for $J/g\gg1$, as expected for a Bose-Einstein condensate. Note, however, that this limit is only approached very gradually and for $N>1$, there is a large intermediate region with partially bound and partially unbound photonic states. 

As the distance between the TLA is increased, the picture remains qualitatively the same. Interestingly, however, the phase transition gets sharper for larger $d$ and occurs at a much lower coupling strength. In particular, at larger densities, the critical coupling strength $g_c$ is reduced by more than an order of magnitude by going from $d=1$ to $d=2$. We attribute this trend to the exponential localization of the bound-photon, such that it requires much higher hopping amplitudes for neighboring bound-states to start to hybridize significantly~\cite{calajo_atom-field_2016}. Note that a superfluid-Mott transition between separated atom-photon bound states has also been discussed in Ref.~\cite{shi_effective_2018}. In general, the critical coupling strength~$g_c$ for the transition can be estimated by a mean-field approach at the level of unit-cells, as explained in Appendix~\ref{app:mean_field}. These estimates are represented by the vertical dashed lines in Fig.~\ref{fig:var_sweep} and accurately predict the transition point.

\subsection{The Mott-superfluid phase transition in waveguides with interacting photons}

We now study the effect of the photon-photon interaction on the Mott-Superfluid phase transitions. In contrast to the case $U=0$ discussed above, we obtain qualitatively different phase diagrams for scenarios with $d=1$ (dense impurities) and $d>1$ (dilute impurities). This difference arise from the possibility of photons becoming detached from the TLAs and occupying the empty sites between adjacent impurities when $d>1$. In the following, we address these cases separately. 

\subsubsection{Dense impurities}

For $d=1$ and $U=0$, the system reduces to the JCH model discussed above. When an additional repulsive photon-photon interaction is added, both the JC-type nonlinearity as well as the Kerr interaction can drive the system between the superfluid and the Mott insulating phase. Indeed, for unit filling with $N=1$, exact numerical simulations show that both types of nonlinearities, as well as their combined presence, lead to a qualitatively similar behavior.

In contrast, for $N\geq 2$ excitations per unit cell, the photons and atoms can arrange in different types of Mott insulating states, which depend on the relative strength of $U$ and $g$. For strong coupling $g$, the ground state of each unit cell is determined by the JC interaction, which favors an equal superposition between the excited and unexcited TLA. Thus, for $N=2$, the resulting Mott insulator state is expected to be of the form
\begin{equation}
    \ket{\psi}_{\rm MI}^{\rm I} \approx  \frac{1}{2^{L/2}}\prod_x  \left( \ket{e,1} - \ket{g,2} \right)_x.
\end{equation}
In the opposite limit, $U\gg g$, it is energetically more favorable to distribute the excitations between the TLA and the cavity mode and form an insulating state of the type 
\begin{equation}
    \ket{\psi}_{\rm MI}^{\rm II} \approx  \prod_x   \ket{e,1}_x.
\end{equation}
To distinguish these two Mott insulator states and separate them from the superfluid phase, we introduce the two fluctuation parameters
\begin{align}
\mathcal{V}_{\rm pol}&=\frac{1}{N_{\rm a}}\sum_{x \in A}\left[\langle(b_{x}^\dagger b_{x})^2\rangle-\langle b_x^\dagger b_x\rangle^2\right],
\end{align}
and 
\begin{align}
\mathcal{V}_{\rm atom}&=\frac{1}{N_{\rm a}}\sum_{x \in A}\left[\langle(\sigma^+_x\sigma^-_x)^2\rangle-\langle \sigma^+_x\sigma^-_x\rangle^2\right].
\end{align}
These parameters quantify the fluctuations of the polariton number and the atomic excitation number, respectively.

\begin{figure}
\centering
\includegraphics[width= \linewidth]{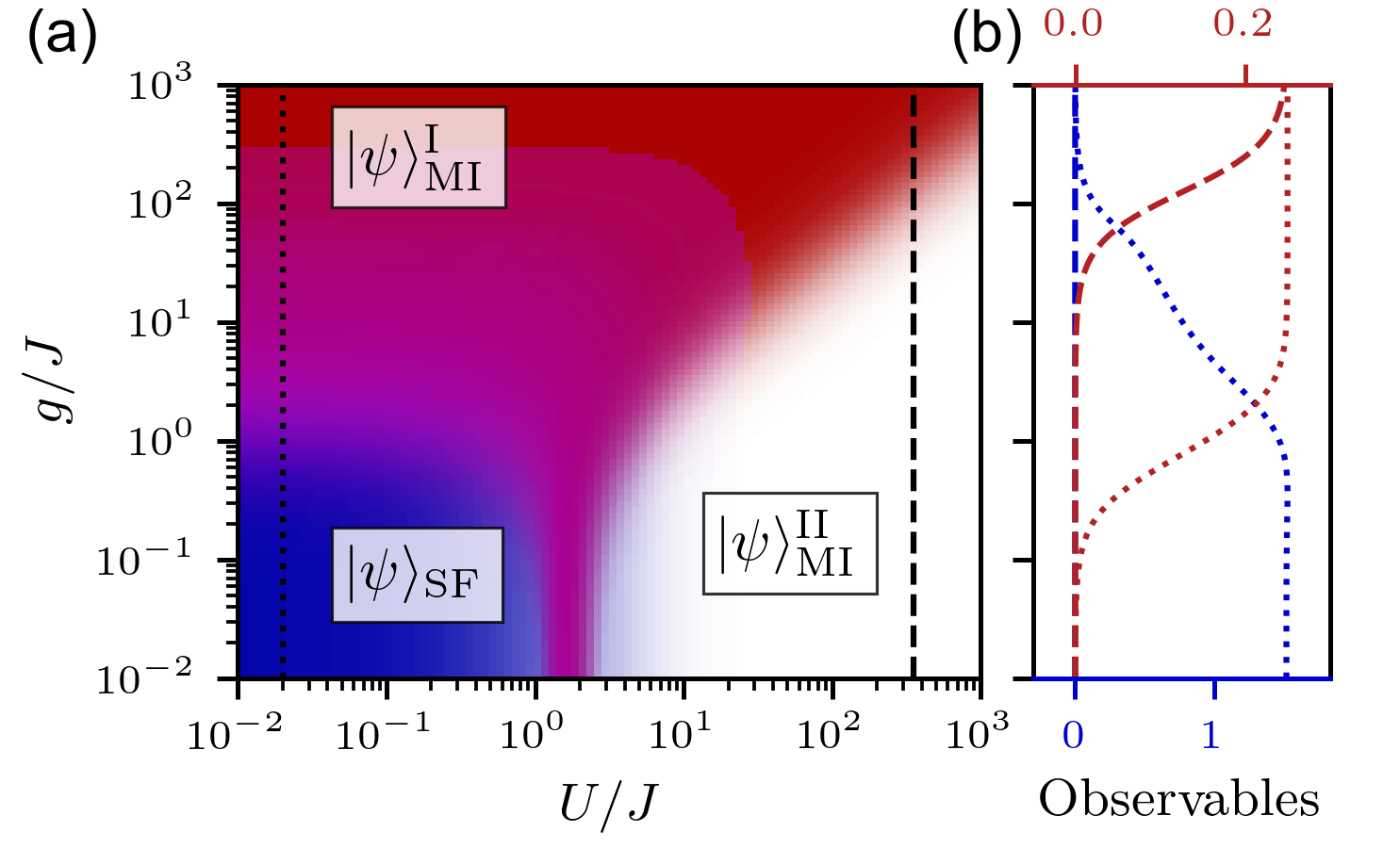}
\caption{Ground-state phase diagram of the interacting waveguide model for $d=1$ and $N=2$ excitations per unit cell. The plot shows the values of the fluctuation parameters $\mathcal{V}_{\rm pol}$ (blue) and $\mathcal{V}_{\rm atom}$ (red) as a function of $g$ and $U$. The same quantities are shown on the right as a function of $g/J$ for the values of $U/J=0.02$ and $U/J=351$ indicated by the dotted and the dashed lines. These results have been obtained from tensor-network simulations with $40$ unit cells and a bond dimension of $\chi=200$. For better visualization of the $\mathcal{V}_{\rm atom}=0$ region, the color contrast between the red and purple region has been increased.
}
\label{fig:JCH_g2}
\end{figure}

In Fig.~\ref{fig:JCH_g2} we plot the ground-state phase diagram for a lattice with $d=1$ and $N=2$ as a function of $g$ and $U$. The colors in these plots indicate the values of  $\mathcal{V}_{\rm pol}$ (blue) and $\mathcal{V}_{\rm atom}$ (red), respectively. The fluctuations of the polariton number vanish for $g\gg J$ and $U\gg J$ in both Mott-insulating states, $\ket{\psi}_{\rm MI}^{\rm I}$ and $\ket{\psi}_{\rm MI}^{\rm II}$. Thus, a nonvanishing $\mathcal{V}_{\rm pol}>0$ indicates a superfluid phase. Instead, the fluctuations of the atom number vanish only in the state $\ket{\psi}_{\rm MI}^{\rm II}$, while they are maximized in the state $\ket{\psi}_{\rm MI}^{\rm I}$. In the phase diagram in Fig.~\ref{fig:JCH_g2}, we thus observe a transition between the two Mott insulating states at sufficiently high values of $U$ and $g$. Note, however, that this is not a sharp transition and follows approximately the gradual variation of the mixing angle $\theta_{n=2}$.

Note that in Fig.~\ref{fig:JCH_g2}, the purple region, where both type of fluctuations are present, $\mathcal{V}_{\rm pol}>0$ and $\mathcal{V}_{\rm atom}>0$, indicates a superfluid regime, where photons can still propagate through the waveguide, but some of the photons are already bound to the atomic impurities. Interestingly, this phase extends to arbitrarily small values of $g$ and separates the regular, photonic superfluid phase from the Mott-II phase. The enhanced fluctuations of the atomic population can thus be used as an alternative way to detect this transition.

\subsubsection{Dilute impurities}
\begin{figure*}
\centering
\includegraphics[width = \linewidth]{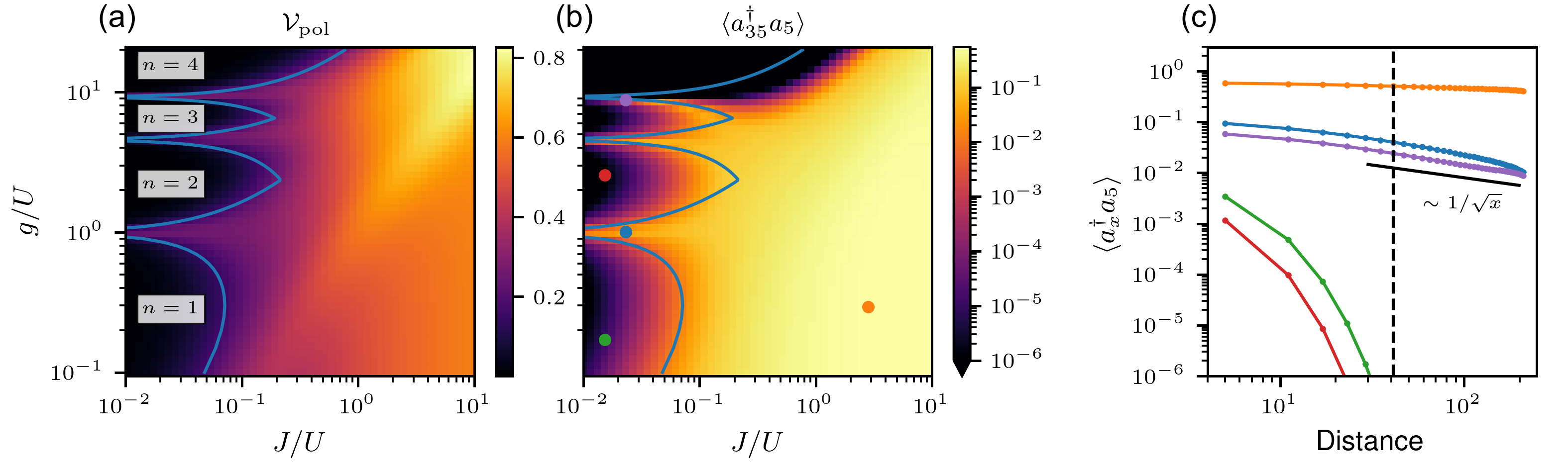}
\caption{Ground-state phase diagram in the dilute impurity limit for a lattice with $d=6$ and $N=4$ excitations per unit cell. (a) Plot of the variance of the polariton number, $\mathcal{V}_{\rm pol}$, as a function of $g$ and $U$. (b) Plot of the two-point correlations $\langle a^{\dagger}_{x} a_{y} \rangle$ evaluated for fixed sites $x=35$ and $y=5$, which are located in between impurities and are $6$ unit cells apart.
In (a) and (b), the blue lines represent the transition line between the superfluid and the Mott phases obtained from the minimum of $J_{n\rightarrow n-1}$ and $J_{n+1\rightarrow n}$ given in Eq.~\eqref{eq:Jcrt_transition_energy}.
(c) Dependence of the two-point correlations $\langle a^{\dagger}_{x} a_{0} \rangle$ on the distance $x$ for different parameters indicated by the colored dots in (b). This plot shows an algebraic scaling $\sim 1/\sqrt{x}$ for the superfluid stripes in between two Mott lobes and an exponential scaling inside the Mott insulating regions.
Simulations where performed for a total of 40 unit cells and bond dimension of $\chi=200$.
}
\label{fig:multi_phase_sweep}
\end{figure*}

We now consider a spacing between atoms that is larger than the number of photons per unit cell. In this setup, the same parameter sweep gives a strikingly different picture, which is shown in Fig.~\ref{fig:multi_phase_sweep} for $d=6$ and $N=4$ excitations per unit cell. Instead of a single superfluid-Mott transition, we now observe multiple transitions and the appearance of distinct Mott lobes. These features are familiar from the phase diagram of the BH model~\cite{fisher1989BosonLocalizationSuperfluidinsulator} or the JCH model for $d=1$~\cite{angelakis2017QuantumSimulationsPhotons}. 
We emphasize though that the phases of these other models are usually studied as a function of the chemical potential $\mu$, while here the Mott lobes are established at a fixed density as a function of the atom-photon coupling $g$. The different Mott lobes represent states with a different, but well-defined number $n$ of photons bound to each atomic impurity, as seen in the plot of the 
polariton fluctuations $\mathcal{V}_{\rm pol}$ in Fig.~\ref{fig:multi_phase_sweep}(a). The fixed total number of excitations also implies that there is a maximal number of Mott lobes (in the current example $n=N=4$). For other lobes with $n<4$, the remaining unbound photons are confined to regions in between neighboring atoms, where they are unable to carry long-range correlations. This situation is similar to the ground state configuration shown in Fig.~\ref{fig:localized_photons} for the case of two impurities and $U/g=1.4$. In the extended waveguide, this behavior translates into exponentially decaying photonic correlations (or photonic coherence) $\langle a_x^\dag a_{x+r}\rangle$, which are plotted in Fig.~\ref{fig:multi_phase_sweep}(b) for a fixed distance and in  Fig.~\ref{fig:multi_phase_sweep}(c) as a function of $r$. Within the lower Mott lobes, this rapid decay of correlation occurs even though there are unbound photons present in the lattice. 

For sufficiently large $J$, all the Mott insulating states melt into a superfluid phase, which is characterized by long-range photonic correlations. In the limit $J\rightarrow\infty$ this superfluid phase is a Bose-Einstein condensate of photons, which means that almost all photons occupy the lowest momentum mode and $\langle a_x^\dag a_{x+r}\rangle\simeq N/d$. A slightly smaller value is observed for the corresponding curve in Fig.~\ref{fig:multi_phase_sweep}(c) due to the finite $J/U\simeq3$ considered. For the maximal Mott lobe ($n=4$) the transition into the superfluid phase occurs through a detachment of the bound photons at rather high values of $U/J$. For the lower lobes, this transition is additionally driven by the unbound photons, which eventually overcome the effective barriers imposed by the impurities. A similar mechanism also occurs in between the Mott-lobes, where narrow regions with long-range photonic correlations are observed. At these values, the situation is similar to the case $U/g=1.2$ in Fig.~\ref{fig:localized_photons}: the impurities become transparent but the unbound photons are still strongly interacting. Therefore, within these narrow regions between the Mott lobes, the systems behaves as a dilute gas of hard-core bosons, for which long-range, but algebraically decaying correlations, $\langle a_x^\dag a_{x+r}\rangle\sim 1/\sqrt{|r|}$, are expected \cite{giamarchi}. Such a dependence is indeed observed in our numerical simulations [see Fig.~\ref{fig:multi_phase_sweep}(c)].    

In the limit $J\rightarrow0$, the Mott-lobes extend over the region $g_{\rm b}(n)\leq g \leq g_{\rm b}(n+1)$ and also the transitions between the lobes can be understood from the effective single-particle model developed in Sec.~\ref{sec:Groundstate_transition}. This model focuses on the transition region between the lobes with $n$ and $n-1$ bound photons, where the impurity sites are approximated by an on-site energy offset $\epsilon_n$ and a hopping impurity~$\alpha_n$ for the unbound photons. From this model, a rough estimate for the transition into the superfluid phase can be obtained by evaluating the point above which the unbound photons can pass through the energy barrier. This occurs approximately at a critical hopping of
\begin{align}\label{eq:Jcrt_transition_energy}
    J_{n} = \frac{\epsilon_n}{2\alpha_n}.
\end{align}
This value of $J_n$ for different $n$ is represented in blue in Fig.~\ref{fig:multi_phase_sweep}. Although it is based on a very rough estimate, it captures very well the transition between the Mott and the superfluid phases, in particular when characterized in terms of the polariton fluctuations shown in Fig.~\ref{fig:multi_phase_sweep}(a).

\section{Implementations}\label{sec:Implementations}

Hamiltonian~\eqref{eq:Hamiltonian} represents a generic toy model for waveguide QED systems, where both strong emitter-photon couplings as well as strong intrinsic Kerr nonlinearities are present. However, to study the impurity physics discussed in this work, both $g$ and $U$ must be comparable to the tunneling amplitude $J$ and exceed as well the characteristic rate $\Gamma={\rm max}\{\Gamma_{\rm a}, \Gamma_{\rm ph}\}$, at which the TLAs or the photons decay. These conditions are still very challenging to achieve in photonic waveguides with emitters in the optical regime. Therefore, in the following we outline two alternative settings where the predicted phenomena can be already be probed with state-of-the-art experimental techniques.    

\subsection{Superconducting circuits}
\begin{figure}
\centering
\includegraphics[width=\columnwidth]{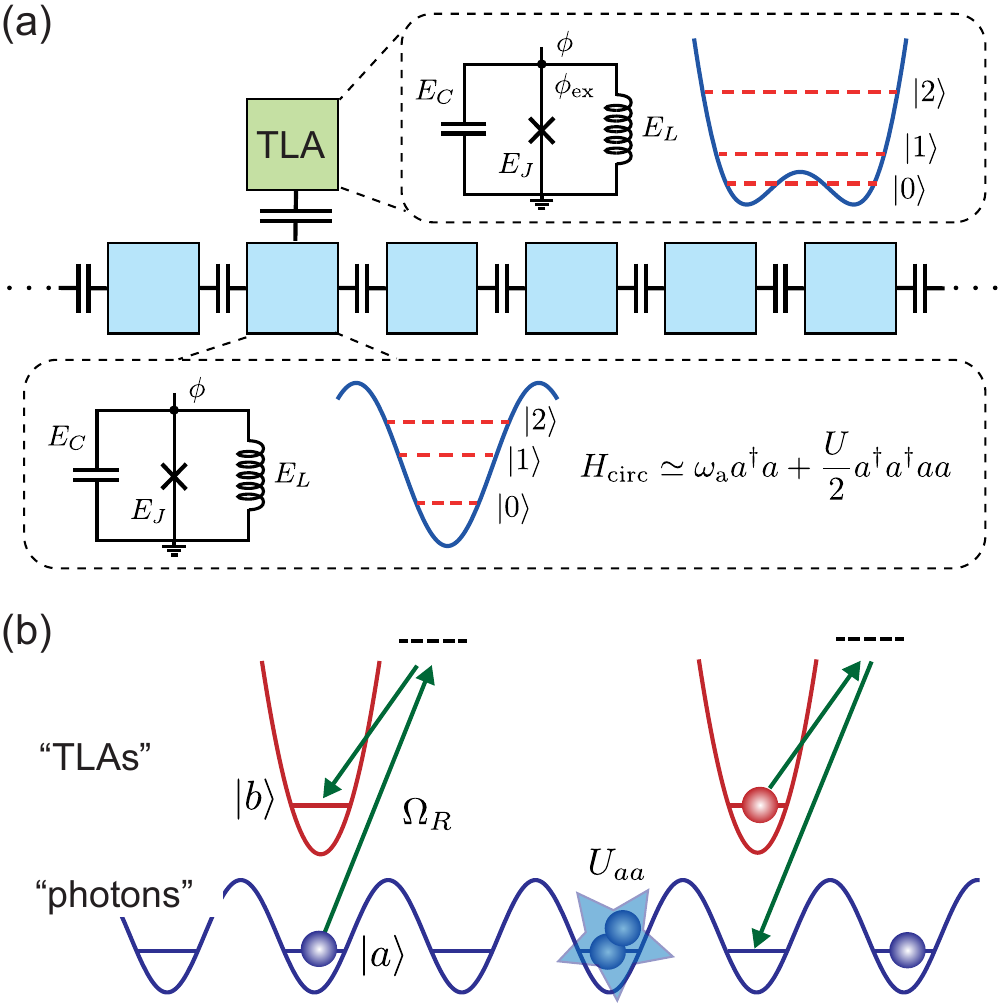}
\caption{Implementations of interacting waveguide QED models. (a) Sketch of a circuit QED lattice with capacitatively coupled, nonlinear $LC$ resonators. Depending on the degree of nonlinearity, the resonators represent either interacting photonic modes (blue boxes) or TLAs (gree box). (b) Realization of an effective waveguide QED system using cold atoms in state-dependent optical potentials. See text for more details.}
     \label{fig:Implementations}
\end{figure}

As already highlighted in the introduction, a natural platform  to study arrays of strongly interacting photons and TLAs are quantized superconducting circuits. In this case, both the TLAs as well as the photonic lattice sites can be represented by lumped-element $LC$ resonators with integrated Josephson junctions to induce nonlinearities~\cite{krantz_quantum_2019}. In terms of the dimensionless phase and charge operators, $\phi$ and $q$, where $[\phi,q]=i$, the Hamiltonian of each of these circuit units is given by  
\begin{equation}
H_{\rm circ} = 4E_C q^2  + \frac{E_L}{2} \phi^2  - E_J \cos(\phi-\phi_{\rm ex}).
\end{equation}
Here, $E_C$, $E_L$ and $E_J$ are the capacitive, inductive and Josephson energy, respectively, and $\phi_{\rm ex}$ is the externally applied magnetic flux through the loop. Depending on the choice of these energies and the value of $\phi_{\rm ext}$, the spectrum of $H_{\rm circ}$ exhibits different signs and degrees of nonlinearities, which can be used to model weakly interacting photons as well as energetically well-isolated TLAs. Further, both the tunnel coupling and the JC interaction can be realized through additional capacitive couplings between the resonator elements. This general approach is sketched in Fig.~\ref{fig:Implementations}(a) and has already been analyzed in the past for implementing BH~\cite{hartmann_quantum_2016,yan2019StronglyCorrelatedQuantum,ma2019DissipativelyStabilizedMott,saxberg2022DisorderassistedAssemblyStrongly,braumuller2022ProbingQuantumInformation,karamlou2024ProbingEntanglement2D}, JCH~\cite{noh_quantum_2017,angelakis2017QuantumSimulationsPhotons,fitzpatrick2017ObservationDissipativePhase,zhang2023SuperconductingQuantumSimulator} and interacting waveguide QED~\cite{wang_supercorrelated_2020} models.

\subsubsection{Transmon circuits}
By setting $\phi_{\rm ex}=0$ and assuming $E_C\lesssim (E_L+E_J)$, we can expand the Josephson energy up to order $\phi^4$ and express the phase and charge operators in terms bosonic annihilation and creation operators, $\phi=\left(\frac{2E_C}{E_J+E_L}\right)^{1/4}(a^\dagger+a)$, $q=\frac{i}{2}\left(\frac{E_J+E_L}{2E_C}\right)^{1/4}(a^\dagger-a)$. We obtain~\cite{blais2021CircuitQuantumElectrodynamics}  
\begin{equation}\label{eq:Transmon}
H_{\rm circ} \simeq \omega_{\rm a} a^\dag a    +  \frac{U}{2}a^\dag a^\dag a a,
\end{equation}
which corresponds to the transmon circuit most commonly used in experiments today. By capacitively coupling many of these nonlinear oscillator, we obtain the Hamiltonian of an interacting cavity array, i.e., the first line of Eq.~\eqref{eq:Hamiltonian}. For typical experimental parameters, $|U|/h\sim 1-100$~MHz and $J/h\sim 10$~MHz~\cite{blais2021CircuitQuantumElectrodynamics,krantz_quantum_2019,Houck2012} we can access both the superfluid and the Mott insulating phases while still operating on timescales that are fast compared to the typical decay rates of $\Gamma/(2\pi)\approx 1-10$ kHz of such circuits~\cite{krantz_quantum_2019}. The two level atoms can be realized either by using a transmon circuit with larger $U$, or a more nonlinear circuit, such as the fluxonium circuit discussed below. 

While experimentally most convenient, an apart problem of the transmon circuit is that the Kerr interaction in Eq.~\eqref{eq:Transmon} is attractive, i.e., $U<0$, in contrast to the repulsive interaction assumed in our analysis. This, however, is not an issue when we work with an isolated system at a fixed total number of excitation and we prepared states adiabatically. Under such conditions, we can apply a unitary transformation $a_x\rightarrow (-1)^x a_x$ to flip the sign of the tunneling term and realize the Hamiltonian $-H$ instead of $H$. For this model, we can then adiabatically prepare the state with the highest energy, which is identical to the ground state of the original Hamiltonian. Note that a similar strategy has been used to study the inverted BH model with cold atoms in optical lattices~\cite{Braun_NegativeTemperature2013}.

\subsubsection{Fluxonium circuits}
As an alternative approach, the waveguide can also be formed by other types of nonlinear circuits. For example, by choosing $\phi_{\rm ex}=\pi$, we obtain a fluxonium circuit with an inverted double-well potential, as sketched in Fig.~\ref{fig:Implementations}(a), and Hamiltonian 
\begin{equation}
H_{\text{circ}}=\sum_{n} (\omega_c n+U_n)|n\rangle\langle n|.
\end{equation}
In this case, the nonlinearities $U_n$ have no simple dependence on the excitation number $n=0,1,2,3,\dots$, but in general $U_n>0$ for the few lowest eigenstates $|n\rangle$. In this approach, some of the details of the bound states and the ground-state phase diagrams change, but most of the analysis presented in this work can be readily adapted to account for such a general local interaction. 
For example, the critical coupling $g_{n}$ of the transition between two neigboring Mott lobes is a solution of the polariton equation
\begin{align}\label{eq:critical_g_def}
    E_{n,\pm}(g_n)=E_{n-1,\pm}(g_n)+\omega_c,
\end{align}
where on resonance the polariton energies are given by
\begin{align}
    E_{n,\pm}=\omega_c n + \frac{U_n + U_{n-1}}{2}\pm\frac{1}{2}\sqrt{(U_n - U_{n-1})^2+4ng^2}.
\end{align}
A generic nonlinearity can still lead to solutions to Eq.~\eqref{eq:critical_g_def} and thus to transitions between different Mott lobes when varying the coupling~$g$.
Therefore, the use of fluxonium circuits is a possible way to implement waveguide models with repulsive photon-photon interactions, although the phase boundaries must be reevaluated.

\subsection{Cold atoms}
Another interesting platform for implementing interacting waveguide QED systems are so-called matter-wave polaritons in optical lattices \cite{krinner_spontaneous_2018, kwon_formation_2022, kim_super_2025}. 
Here, both the TLAs as well as the photons are represented by two different internal states $|a\rangle$ and $|b\rangle$ of cold bosonic atoms. As depicted in Fig.~\ref{fig:Implementations}(b), with the help of state-dependent optical trapping techniques, atoms in state $|b\rangle$ can be tightly trapped by an optical tweezer at a fixed  location in space and represent a stationary TLA in our waveguide model. Atoms in the other state $|a\rangle$ only experience a shallow optical lattice potential along which they can freely propagate. By coupling both internal states via a two-photon Raman transition, a coherent conversion between the atom-like state $|b\rangle$ and the photon-like state $|a\rangle$ can be induced, mimicking a JC interaction~\cite{deVegaPRL2008,Navarrete-Benlloch_2011}.    

Following the usual derivation~\cite{Jaksch1998,Rabl2003}, the Hamiltonian for this cold atom system can be written as 
\begin{equation}\label{eq:Hatom}
\begin{split}
    H_{\rm atom}   = \,\, &  \sum_{x\in A} \delta_R  b^{\dagger}_x b_{x} + \frac{\Omega_R}{2}  \left(a^{\dagger}_{x} b_x + a_{x} b^\dag_x\right) \\
    &-  J \sum_x  \left(a^{\dagger}_x a_{x+1} + {\rm H.c.} \right)+H_{U}(a_x,b_x),
\end{split}
\end{equation}
where $a_x$ ($b_x$) is the bosonic annihilation operator that creates an atom in lattice site $x$ in state $|a\rangle$ ($|b\rangle$). Here, $\delta_R$ and $\Omega_R$ denote the detuning and the Rabi-frequency of the Raman transition, respectively, and we can identify $g\equiv \Omega_R/2$ in our waveguide model. At each site, the atoms interact according to  
\begin{align}
H_{U}(a,b)= \frac{U_{a}}{2} a^\dag a^\dag a a + \frac{U_{b}}{2} b^\dag b^\dag b b + U_{ab} a^\dag a b^\dag b,
\end{align}
where the $U_{a}$, $U_{b}$, and $U_{ab}$ depend on the state-dependent scattering lengths and can be tuned, for example, using Feshbach resonances. To realize the interacting waveguide model, we are interested in conditions $U_{ab}\rightarrow 0$ and $U_a\sim J\sim g \ll U_b$, such that at most one atom in state $|b\rangle$ can occupy one of the tweezer sites. In addition, the optical lattice potential and the tweezer can be slightly displaced to reduce the spatial overlap of the wavefunction and thereby any residual $U_{ab}$. We emphasize that experimental techniques to study the BH type physics~\cite{ColdAtom_Review} as well as effective light-matter interactions~\cite{krinner_spontaneous_2018} are already well-established in this platform.

\section{Conclusion}\label{sec:Conclusions}
In summary we have analyzed the few- and many-body ground-states of a waveguide QED system with atomic impurities and interacting photons. Specifically, we have studied the binding of photons to atomic impurities and showed how the interplay between the JC attraction and the photon-photon repulsion affects the possible configurations of bound states as well as the nature of correlations that are established across multiple impurity sites. Based on large-scale numerical simulations, we have shown how these basic processes lead to a rich ground state phase diagram for this model with different types of insulating and superfluid phases.  

While the investigate model is representative of a large class of interacting waveguide QED systems, we have argued that the described physics can already be investigated with state-of-the-art circuit QED and cold-atom setups. These systems are very promising for quantum simulation applications and in this broader context, an interesting observation of our analysis is the appearance of Mott lobes as a function of the JC-interaction strength. This coupling effectively provides a tuning knob to controls the capture or release of photons and thus the density of the remaining photons in the waveguide. The same concept could be applied also in other photonic quantum simulation schemes to mimic or locally control chemical potentials.

\begin{acknowledgments}
We thank Enrico Di Benedetto, Francesco Ciccarello, Federico Roy and Dominik Schneble for many stimulating discussions. We acknowledge support by the Deutsche Forschungsgemeinschaft (German Research Foundation)–522216022. This research is part of the Munich Quantum Valley, which is supported by the Bavarian state government with funds from the Hightech Agenda Bayern Plus.
\end{acknowledgments}

\appendix

\section{Numerical Methods}\label{app:Numerics}
For our numerical simulations, we primarily use two methods, exact diagonalization and Matrix Product State (MPS) methods \cite{schollwock_matrix_2013, jose_garcia-ripoll_time_2006}. In the latter case, and since we exclusively consider 1D systems, we used MPS methods from the python package \textbf{TeNPy} \cite{hauschild_tensor_2024, hauschild_efficient_2018} and the Density Matrix Renormalization Group (DMRG) algorithm \cite{schollwock_density-matrix_2011} to compute the ground state. We simulate large but finite systems of about 40 unit-cell with open boundary conditions. At this size, we find no noticeable effect of the boundary on the bulk physics.

\section{Mean-field analysis}\label{app:mean_field}

The mean-field analysis has been widely used in the literature to predict the Mott-superfluid transition in the BH model~\cite{fisher1989BosonLocalizationSuperfluidinsulator} and the JCH model~\cite{greentree_quantum_2006,koch2009SuperfluidMottInsulator}. In both cases the mean-field analysis has been applied to a unit cell of size $d=1$. Here we extend this analysis to periodic lattices with an arbitrary lattices spacings, and use it to predict the Mott-superfluid transitions shown in Fig.~\ref{fig:var_sweep}.

\begin{figure}
    \centering
    \includegraphics[width=\linewidth]{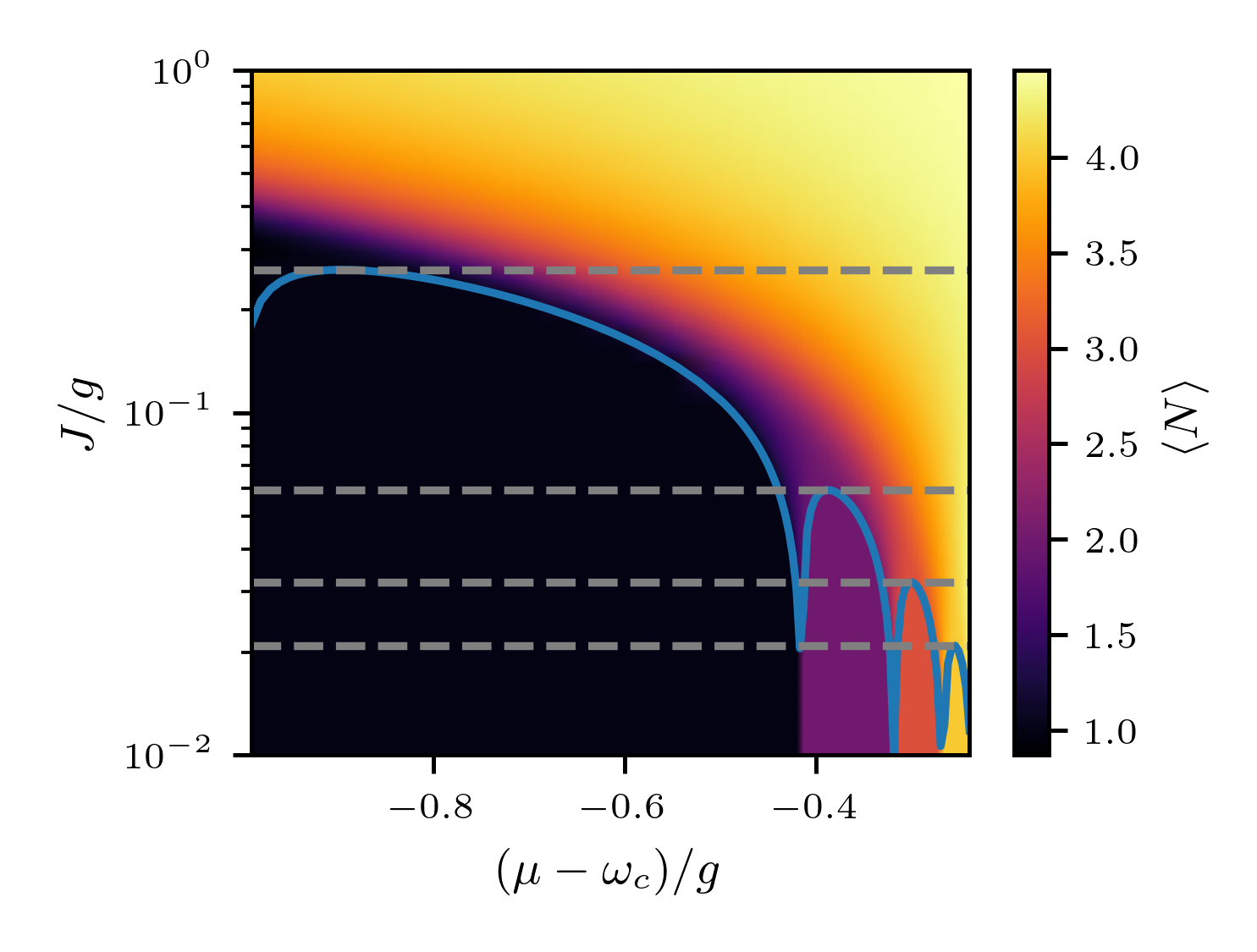}
    \caption{Mean-field analysis for $U=0$ and $d=3$ cavities per unit-cell. In blue: the critical hopping obtained by solving Eq.~\eqref{eq:MF_transition_eq}. Density plot: mean number of excitation per unit-cell in the mean-field ground state, showing the Mott lobes when varying the chemical potential. Grey dashed line: maximum critical hopping in each Mott lobe.}
    \label{fig:app:MF}
\end{figure}

In the mean-field analysis, we assume that the wavefunction is factorized between every unit cell $\ket{\Psi}=\otimes_{p}\ket{\psi_p}$, where $p$ is the unit-cell index.
The Hamiltonian governing each unit-cell wavefunction $\ket{\psi_p}$ is obtained from the substitution $ab\rightarrow \langle a\rangle b + a \langle b\rangle - \langle a\rangle \langle b \rangle$ for each pair of operators $a,b$ belonging to different unit-cells.
We denote by $a_{p,j}$ the $j$-th annihilation operator of the $p$-th unit-cell, and we consider the two-level system coupled to $a_{p,1}$.
Assuming the same average value $\phi_1=\bra{\psi_p}a_{p,1}\ket{\psi_p}$ and $\phi_2=\bra{\psi_p}a_{p,d}\ket{\psi_p}$ for every unit-cell~$p$, we obtain the mean-field Hamiltonian $H_\MF$ of each unit-cell, which depends on the two order parameters $\phi_1$ and $\phi_2$. It is given by 
\begin{align}
    H_{\MF}
    =& 
    g(\sigma^- a_1^\dagger + \sigma^+ a_1) + 
    \sum_{j=1}^d \frac{U}{2}a_j^\dagger a_j^\dagger a_ja_j \nonumber\\
    &- J\sum_{j=1}^{d-1}(a_{j+1}^\dagger a_j + {\rm H.c.}) - J(\phi_1 a_d^\dagger + \phi_2 a_1^\dagger + {\rm H.c.}) \nonumber\\
    &+ J(\phi_1^*\phi_2 + {\rm c.c.}) - 
    \mu (\sigma^+\sigma^- +\sum_{j=1}^d a_j^\dagger a_j),
\end{align}
where we omit the unit-cell index $p$, and where $\mu$ is an effective chemical potential corresponding to a Lagrange multiplier to be tuned to fix the average number of excitations per unit-cell. 
The mean-field ground-state $\ket{\psi_g}$ is obtained by minimizing the ground state energy $E_g(\phi_1,\phi_2)$ of $H_\MF$ with respect to $(\phi_1,\phi_2)$, under the self-consistency constraints $\phi_1=\bra{\psi_g}a_{1}\ket{\psi_g}$ and $\phi_2=\bra{\psi_g}a_{d}\ket{\psi_g}$.
The mean-field parameters $\phi_1$ and $\phi_2$ can be chosen real by $U(1)$ symmetry.
We then expand
 \begin{equation}\label{eq:MF_energy_expansion}
     E_g(\phi_1,\phi_2) = E_0 + u_1 \phi_1^2 + u_2 \phi_2^2 + 2v \phi_1\phi_2 + \mathcal{O}(\phi^4),
 \end{equation}
where the odd orders in $\phi_1,\phi_2$ vanish due to the $U(1)$ symmetry.
For small hopping $J$ (in the Mott regime), the point $\phi_1=\phi_2=0$ is a saddle point of $E_g(\phi_1,\phi_2)$, and it is the only point where the self-consistency is satisfied. 
For large hopping $J$ (in the superfluid regime), the point $\phi_1=\phi_2=0$ turns to a local maxima, with two lower energy points as saddle points for non-zero order parameter $\phi_1,\phi_2\neq 0$.
This second-order phase transition corresponds to a vanishing Hessian determinant in Eq.~\eqref{eq:MF_energy_expansion}, 
\begin{equation}\label{eq:MF_transition_eq}
    u_1u_2 - v^2 = 0.
\end{equation}
The parameters $u_1,u_2,v$  can be obtained semi-analytically by diagonalizing the $\phi_1=\phi_2=0$ Hamiltonian
\begin{align}
    H_\MF(\phi_1=0,\phi_2=0) = \sum_k E_k \ket{k}\bra{k},
\end{align}
such that, using second-order perturbation theory,
\begin{align}
    u_1 &= J^2 \sum_{k\neq 0}\frac{\left|\bra{k}(a_d+a_d^\dagger)\ket{0}\right|^2}{E_0 - E_k}, \\
    u_2 &= J^2 \sum_{k\neq 0}\frac{\left|\bra{k}(a_1+a_1^\dagger)\ket{0}\right|^2}{E_0 - E_k}, \\
    v &= J+ J^2 \sum_{k\neq 0}\frac{\textrm{Re}~\bra{0}(a_1+a_1^\dagger)\ket{k}\bra{k}(a_d+a_d^\dagger)\ket{0}}{E_0 - E_k}.
\end{align}
The critical hopping~$J_c$ can then be found as the root of Eq.~\eqref{eq:MF_transition_eq}. Note that the Hamiltonian for $\phi_1=\phi_2=0$  still depends on $J$ for $d>1$, in contrast to the JCH model~\cite{greentree_quantum_2006,koch2009SuperfluidMottInsulator}. As a result, Eq.~\eqref{eq:MF_transition_eq} is a highly non-linear equation in $J$.
We thus find the solution $J_c$ numerically using a root-finding algorithm.
It is represented in the blue curve in Fig.~\ref{fig:app:MF} as a function of the chemical potential~$\mu$ for $U=0$ and $d=3$ cavities per unit-cell.
The average number of excitations in the mean-field ground state is represented in a density plot, which shows that the chemical potential changes the average number of excitations per unit cell.

To compare to the DMRG result of Fig.~\ref{fig:var_sweep}, where the total number of excitations $N$ per unit-cell is fixed, we choose the chemical potential $\mu$ giving the largest $J_c$ in the corresponding Mott lobe (gray dashed lines in Fig.~\ref{fig:app:MF} for $N=1,2,3,4$).
The corresponding~$g/J_c$ matches the point of phase transition of the DMRG result (red dashed lines in Fig.~\ref{fig:var_sweep}).

\newpage

\end{document}